\documentclass[12pt,preprint]{aastex}

\begin{document}

\title{Light Curves of Microlensed Type Ia Supernovae}
\author{Hamed Bagherpour %
\footnote{hamed@nhn.ou.edu %
} , Ronald Kantowski %
\footnote{kantowski@nhn.ou.edu %
} , David Branch %
\footnote{branch@nhn.ou.edu %
} }

\affil{University of Oklahoma, Department of Physics and Astronomy,\\
 Norman, OK 73019, USA }

\author{Dean Richardson %
\footnote{richardson@denison.edu%
} }

\affil{Denison University, Department of Physics and Astronomy, \\
 Granville, OH 43023, USA }

\begin{abstract}
A detailed description of the apparent light curves of microlensed
SNe Ia as extended and expanding sources is presented. 
We show that microlensing amplification 
can have significant effects on supernova observation. A model light curve
is used to compare lensed and unlensed cases and we find that significant
changes in shape can occur because of microlensing. We briefly discuss
the probability of observing such effects as well. Throughout the
paper we consider source and deflector redshifts at 1.0 and 0.05 
respectively, and limit
our work to spherically symmetric deflectors. 
\end{abstract}


\section{Introduction }

Out of a number of distance indicators, supernovae have emerged as
the most promising standard candles. Due to their significant intrinsic
brightness and relative ubiquity they can be observed in the local
and distant universe. Several teams including the High-z Supernova
Search \citep{Schmidt98} and the Supernova Cosmology Project \citep{Perlmutter99}
have been searching for supernovae at higher redshifts
since the early 1990's. Light emitted from these `standard candles'
is subject to lensing by intervening objects while traversing the
large distances involved \citep{KVB95}; the further the
light source, the higher its chance of being significantly lensed.
In fact, for cosmologically distant sources, the probability is high
that a distant point source will be `imaged' \citep{Press73,BK76,Wyithe02},
particularly by stellar-size objects (microlensing). 
While the systematic errors introduced by K-correction, selection 
effects, and possible evolution can be removed, lensing might ultimately 
limit the accuracy of luminosity distance measurements \citep{Perlmutter03}.
Only a large sample of SNe at each redshift can be used to
characterize the lensing distribution and to correct for the effect
of lensing.

Properties of microlensed supernovae have been studied previously.
\citet{Schneider87} presented the time-dependent amplification
of supernovae caused by the expansion of the photosphere, and showed that 
the related polarization of a supernova
is not likely to exceed 1\%. For cosmologically distant supernovae
a 1\% effect is practically impossible to detect. 
\cite{Linder88} 
studied amplification of supernovae and developed approximate formulae for 
the amplification probability distribution. \citet{Rauch91} studied microlensing of SNe Ia
by compact objects and calculated the resulting amplification probability
distributions using Monte Carlo simulations. \citet{Kolatt98}
calculated the light curves of type Ia supernovae microlensed by intracluster 
MACHO's assuming the point-deflector (with shear) model of \citet{Chang84}.   

In this paper, we demonstrate how microlensing by a single stellar-size deflector can affect 
light curves of cosmologically distant SNe Ia. We use the simple 
Schwarzschild lens model to calculate amplifications. 
We ignore the additional amplification caused by the macrolensing  
introduced by the hosting galaxy, and concentrate on the time dependent
effects caused by a single moving stellar-size deflector.  We model the SNe Ia 
by expanding light sources with realistic radial surface intensity profiles.
  
In the next section, we present a model light curve for type Ia supernovae.
$\S$ 3 is devoted to a brief review of the gravitational lensing
effect, and in $\S$ 4 we present the results of our calculations.
Throughout, we assume a flat Friedmann-Lama\^{\i}tre-Robertson-Walker
cosmological model with $\Omega _{m}=$ 0.3, $\Omega _{\Lambda }=$
0.7, and h$_{100}=$ 0.67 to calculate the distances to the source
and deflector.

\section{Light Curves of Type Ia Supernovae }

In general, spectacular supernovae explosions can be interpreted in
terms of two concepts \citep{Arnett96}: A shock wave running through
a stellar envelope, and the radioactive decay of newly synthesized
$^{56}$Ni to $^{56}$Co and then to $^{56}$Fe. The nature of the
explosion is the expansion away from the region of energy release.
The explosion becomes more spherically symmetric with time, trending
toward a small-scale Hubble type flow.

In the meantime, various aspects of the event including shock emergence,
radiative diffusion, heating, and recombination determine the evolution
of the luminosity, i.e., the light curve, which is usually exhibited
as the absolute magnitude as a function of time. In the case of SNe
Ia, it is the shape of the light curve that can be used to measure
cosmological parameters such as Hubble constant; see, for instance,
\citet{Riess95}.

To model the intrinsic Type Ia supernova light curve we used a combination
of two existing analytical models; for the peak of the light curve
(early time) we used the model of \citet{Arnett82} while for the tail
of the light curve (late time) we used the deposition function described
by \citet{Jeffery99}. The basic assumptions made in this model are: a
homologous expansion of the photosphere, radiation pressure dominates
at early times, the diffusion approximation is valid at early times,
the optical opacity is constant for the light curve peak and the gamma-ray
opacity is constant for the tail, $^{56}$Ni is present in the ejecta
and its distribution is peaked toward the center of the ejected mass
at small initial radius. The details of this combined model are in
\citet{Richardson05}. The model parameters have been fixed for
a typical SN Ia: the kinetic energy is 1 foe ($10^{51}$ erg), the
total ejected mass is 1.4 M$_{\odot }$, and the $^{56}$Ni mass has
been set to 0.6 M$_{\odot }$. These values produce a SN Ia with a
peak absolute magnitude of about -19.5.

To test the model we varied the kinetic energy, nickel mass, and the
rise time until we achieved the best $\chi ^{2}$ fit to the observed
data of a few SNe Ia. The model worked well as can be seen in Figure
1, where the fit of SN1990N is shown with E$_{k}=0.60$ foe, M$_{Ni}=0.62$
M$_{\odot }$, t$_{rise}=21$ days, and the reduced $\chi ^{2}=0.43$.
We use this supernova because it was well observed (the light curve
had good coverage) and it is photometrically characteristic of most
SNe Ia.

\section{Microlensing of Extended Sources }

\subsection{Basics }

The linearized Einstein theory for a static gravitational field gives
a bending angle for light rays passing through a weak gravitational
field of 
\begin{equation}
\mbox{\boldmath{$\alpha$}}=-\frac{2}{c}\int _{-\infty }^{+\infty }
\mbox{\boldmath{$\nabla$}}\phi \, dt\: ,\label{eq:1}
\end{equation}
where $\phi$ is the Newtonian gravitational potential satisfying
the boundary conditions $\phi \rightarrow 0$ at infinity, and where
the integral is performed along the light path in the absence of the
gravitational field. For a point deflector the bending angle is simply 
\begin{equation}
\mbox{\boldmath{$\alpha$}}=-\frac{4Gm_{d}}{c^{2}r}\, \hat{r}\: ,\label{eq:2}
\end{equation}
where \mbox{\boldmath{$r$}} is the light ray's impact vector and M$_{d}$
is the deflector's mass. \citet{BKN73}, and
\citet{BK74,BK76} used the 2-component nature of
\mbox{\boldmath{$\alpha$}} to replace it with the complex scattering function $I(z)$,
where $z=x+iy$ is the complex equivalent of the 2-d vector $\mathbf{r}=x\hat{\imath }+y\hat{\jmath }$.
Using $I(z)$, the relation between the source position, $z$, and
image position, $z_{o}$, when both are projected onto the plane of
the deflector (also called the sky plane) is 
\begin{equation}
z=z_{o}-\frac{4GD}{c^{2}}I^{*}(z_{o})\: .
\label{z=z0}
\end{equation}
The scaled (effective) distance $D$ is defined as $D=D_{ds}D_{d}/D_{s}$, where
the deflector-source distance, $D_{ds}$, the observer-deflector distance,
$D_{d}$, and the observer-source distance, $D_{s}$, are all the
same type distances, e.g., apparent size distances.

In the case of a point deflector (on which we focus in this paper),
the scattering function takes a very simple form: 
\begin{equation}
I(z_{o})=\frac{m_{d}}{z_{o}}\: ,\label{eq:4}
\end{equation}
and the equation (\ref{z=z0}) reduces to 
\begin{equation}
z=z_{o}-\frac{r_{E}^{2}}{z_{o}^{*}}\: ,\label{eq:5}
\end{equation}
where $r_{E}\equiv \sqrt{4Gm_{d}D\, c^{-2}}=\sqrt{2r_{S} D}$ is the
Einstein ring radius, and $r_{S}$ is the deflector's Schwarzschild
radius. This equation has two separate solutions (images) for any
source position $r=|z|$, 
\[
r_{p}=\left|z_{p}\right|=\frac{1}{2}\left(\sqrt{r^{2}+4r_{E}^{2}}+r\right)\: ,
\]
and 
\begin{equation}
r_{s}=\left|z_{s}\right|=\frac{1}{2}\left(\sqrt{r^{2}+4r_{E}^{2}}-r\right)\: .\label{eq:6}
\end{equation}
Both images are in line with source and deflector. The `primary'
image, $r_{p}$, lies on the same side of the deflector while the
`secondary' image, $r_{s}$, is on the other side (Fig. 2). In
the case of microlensing, the angular separation of the two images
($\theta _{p}+\theta _{s}$ in Fig. 2) is of the order of micro
arcseconds. Consequently, the two images are seen as a single object
($\S$ 3.2). The Einstein ring occurs when source and deflector are aligned with
the observer $(r=\left|z\right|=0)$ for which $r_{p}=r_{s}=r_{E}$, and due 
to the symmetry of the lensing configuration the image is
actually a ring. For a point deflector, the primary image is outside
the Einstein ring ($r_{p}\geqslant r_{E}$) while the secondary image falls
inside ($r_{s}\leqslant r_{E}$) as can be seen in Figure 3.

The effect of gravitational lensing on the apparent brightness of
a distant source can be computed in various ways. For extended sources
like supernovae it is best to employ the fact that the apparent brightness
is proportional to the image's apparent area, meaning that the brightness
of a source is amplified by a factor $A$:
\begin{equation}
A=\frac{{\cal A}_{o}}{{\cal A}}\: ,
\label{A}
\end{equation}
where ${\cal A}_{o}$ is the area of the image and ${\cal A}$ is the area of the
source both projected on the deflector plane. For a point mass deflector
and a small point-like source the amplification of the primary and
secondary images respectively can be written as 
\[
A_{p}=\frac{r^{2}+2r_{E}^{2}}{2r\sqrt{r^{2}+4r_{E}^{2}}}+\frac{1}{2}\: ,
\]
and 
\begin{equation}
A_{s}=\frac{r^{2}+2r_{E}^{2}}{2r\sqrt{r^{2}+4r_{E}^{2}}}-\frac{1}{2}\: .
\label{Ap&As}
\end{equation}
If the images are unresolvable the combined amplification becomes
\begin{eqnarray}
A & \equiv  & A_{p}+A_{s}=\frac{r^{2}+2r_{E}^{2}}{r\sqrt{r^{2}+4r_{E}^{2}}}\: .
\label{Ap+As}
\end{eqnarray}

If the source is not small, i.e., if $r/r_{E}$ varies significantly
across the source, differential amplification must be taken into account.
Measuring this amplification from a single observation is not possible
since it is practically impossible to figure out the original source
flux. However, if the luminosity of the source varies with time in
a predictable way as with SNe or if the source is of constant brightness
and the lens is moving with respect to the line of sight to the source
[as with observation of bulge stars; see, for instance, \citet{Sumi04}],
the amplification will change with time in a predictable way
and it is possible to determine the amplification. Figure 4 shows
amplification curves of a point source lensed by a spherically symmetric
deflector with mass $m_{d}=1$ M$_{\odot}$.

The relative time delay of the two images (which is due to both the
geometrical path difference of the two light signals and the difference
in the gravitational potential through which the light rays travel
\citep{CK75} turns out to be only a small fraction of
a second as expected for microlensing. Such small time delays between
the two images can therefore have no effect on microlensed supernova
light curve when features change on time scales of days or greater.

\subsection{Amplification of an Extended Source}

As implied in the last section, to obtain the total flux received
from an extended source, an integral of intensity across the source
may be required: 
\begin{equation}
A=\frac{\int _{image}I\, d{\cal A}_{o}}{\int _{source}I\, d{\cal A}}\: .\label{eq:10}
\end{equation}
If the surface brightness is constant across the source or if there
is no differential amplification, the net amplification is simply
given by equation (\ref{A}), where ${\cal A}_{o}$ can be the primary, secondary,
or total image area, giving respectively the primary, secondary, or
total amplification. In the case of a circular extended source with
a projected radius of $a$ at a distance $l$ from the center of a
spherically symmetric deflector ($a$ and $l$ are measured in the
deflector plane), the total area of the combined and unresolvable images
is 
\begin{equation}
{\cal A}^{(total)}=\int _{-\frac{\pi }{2}}^{\frac{\pi }{2}}a\left(a+l\, \sin \varphi \right)
\sqrt{1+\frac{4r_{E}^{2}}{l^{2}+a^{2}+2al\, \sin \varphi }}d\varphi \: 
\end{equation}
After a rather long calculation, the total amplification is found
to be 
\begin{equation}
A[a,\, l,\, r_{E}]=\eta \left\{ \mu _{1}K\left(k\right)+\mu _{2}E\left(k\right)+
\mu _{3}\Pi \left(n,\, k\right)\right\} \: ,
\label{Aellip}
\end{equation}
\citep{Witt94,Mao98} where $K$, $E$, and $\Pi $ are respectively
the first, second, and third complete elliptic integral with 
\[
k=\frac{16alr_{E}^{2}}{\left(l+a\right)^{2}\left(\left(l-a\right)^{2}+4r_{E}^{2}\right)}\: ,
\]
\[
n=\frac{4al}{\left(a+l\right)^{2}}\: ,
\] 
and constants 
\[
\eta =\frac{1}{2\pi a^{2}\sqrt{\left(l-a\right)^{2}+4r_{E}^{2}}}\: ,
\]
\[
\mu _{1}=\left(l-a\right)\left(a^{2}-l^{2}-8r_{E}^{2}\right)\: ,
\]
\[
\mu _{2}=\left(l+a\right)\left(\left(l-a\right)^{2}+4r_{E}^{2}\right)\: ,
\]
\begin{equation}
\mu _{3}=\frac{4\left(l-a\right)\left(a^{2}+r_{E}^{2}\right)}{l+a}\: .
\end{equation}
Using equations (\ref{Ap&As}) and (\ref{Ap+As}), equation (\ref{Aellip})gives amplifications for
both the primary and secondary images. Figure 5 is a plot of the amplification
curves of an extended source moving with speed 0.5 AU per day 
(1 AU day$^{-1}\simeq 1.73 \times 10^{3}$ km s$^{-1}$)
with respect to a deflector of mass 1 M$_{\odot }$. To illustrate combined 
effect of relative 
motion and finite size on  amplification,  in Figure 5 we have assumed
a constant surface brightness across the source (no limb darkening)
and a fixed radius of 178 AU, corresponding to that of a supernova with
a maximum atmospheric speed of 30,000 kilometer per second at eighteen
days after the explosion.  In Figure 5, the source approaches 
the deflector
with various impact parameters (closest approach),
producing different amplification curves.

In what follows we use a more realistic supernova model to calculate 
amplification curves. We assume a ring-like structures
for the intensity $I$ of a spherically expanding photosphere like
that of a SN Ia \citep{Hoflich90}. To model the 2-dimensional brightness
profile across the source we assumed that the emissivity of light coming 
from any volume element in the photosphere of a type Ia supernova is 
constant. We found  
\begin{equation}
I(r) = I_{o} \sqrt {1-\left( \frac{r}{r_{ph}} \right)^{2}}\: ,
\end{equation}
where $I_{o}$ is the intensity at the center and $r_{ph}$ is the radius
of the photosphere. This profile is in agreement with one of the commonly used
family of limb-darkening profiles \citep{Allen73,Claret95}.
The assumption that emissivity per volume
in SNe Ia is constant is a reasonable assumption during the nebular phase ($t>150$ days). 
To see whether the obtained intensity
profile is also applicable to photospheric phase ($t<150$ days) we compared 
it to a normalized intensity distribution function (IDF) calculated using 
W7 model \citep{Nomoto84} for SNe type Ia in U and B bands (courtesy of E. Lentz). The result 
of this IDF calculation as well as our intensity profile can be seen in 
Figure 6. The two models show a reasonable match.

We also estimated the characteristic radius of the optical image of a
type Ia supernova as a function of time.  For the photospheric phase,
we assumed homologous expansion (r=vt) and obtained the radius of the
photosphere as a function of time since explosion by multiplying the
speed at the photosphere, as determined empirically by \citet{Branch05}, by the time.  
For the nebular phase, we assumed that the
radius of the iron--group core expands at a constant velocity of 6000
km s$^{-1}$.  A good fit for the speed is an exponential:
\begin{equation} 
v=9.1\, \textrm{e}^{(-\textrm{t}/36.5\, \textrm{days})}+6.0 \: ,
\end{equation}  
in units of $10^{3}\ \textrm{km s}^{-1}$. Figure 7 shows the expansion velocity as well as the radius
as a function of time.

\subsection{Probability }

The relevant quantity in seeing a microlensing event is the optical
depth $\tau $. It is defined as the probability that a point source
(or equivalently the center of an extended source) falls inside the Einstein ring
of some deflector. The brightness of a point source within
$r_{E}$ is amplified by a factor of at least 1.34. For randomly located point deflectors
the optical depth depends on the mass density of the deflectors 
and not on their number density \citep{Press73}.

Typical values of the optical depths for microlensing of nearby luminous
sources are remarkably small. For instance, \citet{Sumi03}
calculated an optical depth $\tau =2.59_{-0.64}^{+0.84}\times 10^{-6}$
toward the Galactic Bulge (GB) in Baade's window for events with time
scales between 0.3 and 200 days. Because of the small value of $\tau $,
millions of stars need to be monitored when searching for microlensing
in areas such as the GB, the Large Magellanic Cloud, or the
Small Magellanic Cloud. The value of $\tau $ is much higher
for cosmologically distant sources. Assuming that the ordinary stellar
populations of galaxies are the dominant causes of microlensing events,
\citet{Wyithe02} concluded that in a flat universe, at least
1\% of high-redshift sources ($z_{s}\geqslant 1$) are microlensed by stars
at any given time. \citet{Zakharov04} estimated that the optical
depth for microlensing caused by deflectors both localized in galaxies
and distributed uniformly, might reach 10\%  for sources at $z_{s}\sim 2$.
If we look at bulge-bulge lensing , \citet{Han03} give a model 
for the bulge from which a value of 
$\tau=0.98\times 10^{-6}$ is computed. They additionally compute  
an effective column density for deflectors $\Sigma_*=$ 2086 M$_{\sun}$pc$^{-2}$ and a 
characteristic source-lens separation $\overline D = 782$\,pc defined by 
\begin{equation}
\tau=\frac{4\pi G}{c^2}\Sigma_*\overline D.
\end{equation}
If we simply look at the bulge of a similar galaxy at $z= 0.05$ we expect 
a similar $\Sigma_*$ but $\overline D$ becomes the distance to the deflector 
$D_d$ and hence increased by a factor  
$\approx 2.7\times 10^5$. This would bring the 
optical depth up to ~27\% when looking through such a galaxy.
At this distance the size of the bulge is $\sim 1$ arcsec and 
clearly resolvable. The downside is one of alignment. What is the chance of 
a host galaxy being appropriately aligned with a foreground galaxy?
The best place to see this effect seems to be the foregrounds of 
dense clusters. Because the typical galaxy is expected to contain a SN\,Ia 
every 10$^3$ yrs, some $3700(1+z)$ alignments would have to be followed 
for a year to see one event.
 
As mentioned above, the amplification of a point source falling inside the
Einstein ring $r_{E}$ is larger than 1.34. The probability of a larger
amplification is proportionally smaller. The probability of having
an amplification larger than $A$ for a given lensing configuration is 
\begin{equation}
p(A)=u_{A}^{2}\, \tau(z_{s}) \: ,
\label{p(A)}
\end{equation} 
where $u_{A}\equiv b_{A}/r_{E}$ is the normalized impact parameter,
resulting in amplification $A$ \citep{Paczynski86a,Paczynski86b} and $\tau(z_{s})$
is the optical depth for a source at redshift $z_{s}$. For a point
source, the result is
\begin{equation}
u_A=\sqrt{2\left(\frac{A}{\sqrt{A^{2}-1}}-1\right)}
\: .\label{u_A}
\end{equation}
In the next section we use (\ref{p(A)}) to reduce the optical depth to probabilities 
appropriate for higher amplifications. 

\section{Microlensed Light Curves }

In this section we use the model light curve of $\S$ 2 and the
microlensing theory of $\S$ 3 to predict the shape of lensed light
curves of type Ia supernovae. We have calculated absolute magnitudes
of the lensed SNe type Ia in the V-band. We assume that the source
is at redshift $z_{s}=1.0$, hence introducing a time dilation factor
of $(z+1)^{-1}=0.5$, and is lensed by a deflector located at redshift
$z_{d}=0.05$. Figure 8 shows the geometry of our model. To calculate
the amplification as a function of time (eq. [\ref{Aellip}]), we need
to know the distance $l$ between the supernova and the deflector,
projected on the plane of the deflector, 
\begin{equation}
l(t)=\sqrt{l_{o}^{2}+vt\left(vt\mp 2\sqrt{l_{o}^{2}-b^{2}}\right)}\: ,\label{eq:24}
\end{equation}    
where the minus sign is used when the supernova source explodes ($t=0$
and $l=l_{o}$) before getting to the point of closest approach, $b$,
and the plus sign when it explodes after. We also need the Einstein
ring radius $r_{E}$ which is determined by the mass of the deflector
$m_{d}$ and the distances of the supernova and the deflector from
the observer.

We plot light curves for various values of parameters $m_{d}$, $b$,
$l_{o}$, and $v$ (the relative speed of source and deflector projected
on the deflector's plane). Our sample deflector masses are $10^{-3}$ M$_{\odot }$,
1 M$_{\odot }$, and 10 M$_{\odot }$. We take the source to move in
the deflector plane for 1,000 days (observer time) with two relative projected speeds
of the source $v$, 0.1\, AU day$^{-1}$ and 0.5 AU day$^{-1}$.

Figures 9 through 17 show the amplification curves (upper panels)
and the light curves (lower panels) of lensed supernova for nine different
sets of input parameters. These interesting cases have been selected from
a large number of configurations. The top panel of each figure shows
the amplification of the primary and secondary images as well as their
sum (total amplification), and the bottom panel includes light curves
of the two images and their sum (the apparent light curve) as well
as the original (unlensed) light curve of the supernova.

We are plotting amplified absolute magnitudes of the supernova in
the V-band, however, due to the redshift of the source, these light
curves would be observed in the I-band. With the source located at
$z_{s}=1$, M$_{V}>-17$ is too dim to be seen. Nonetheless, we include
the complete amplification and light curves to show their trends over
a period of 1,000 days after the supernova explosion.

We can easily match the features in the light curve in each figure
to the corresponding peak of the amplification curve. Figure 9 
shows a huge overall increase in the brightness of the whole
light curve. For this case the SN explosion occurs when the SN is
near minimum impact $l_0=b=0$. The amplification is strongest
around the peak so the largest decrease in M$_{V}$ happens there
(almost 5 magnitudes). If the mass of the deflector were increased 
to 10 M$_{\sun}$ 
the magnitude of the `narrowed' peak would reach M$_{V}\sim -26$
which is comparable to the magnitude of a bright quasar. Such huge
amplifications can cause a bias in observing supernovae, allowing
one to observe more distant supernovae, and as a result, to increase
the depth (volume) of any supernova survey. Notice that this
light curve looks different from those of strongly lensed supernovae
in which, the lensed curve merely has an overall magnitude shift upward 
due to the static magnification introduced
by the microlens' parent galaxy.

In Figures 10 and 11, the lensed light curve hits a plateau before
falling back on the expected trend of a type Ia supernova. 
For some values of the input parameters it is possible to observe
a second peak in the light curve, e.g., see Figures 12 through 17.
The second peak in each of these figures is bright enough to be observed
and could easily be distinguished from the original
(first) peak.
If the impact parameter in Figure 12 is reduced to $b=0$ 
the second peak would actually be larger than the first. 
Figure 17 is another example of a second peak in the
light curve which is bright enough to be seen although it occurs more
than 400 days after the supernova explosion. If the position of the initial 
explosion were increased to  $l_0=0.1$ the still visible second peak 
would occur 2 years after the first. For many cases, microlensing has a 
less dramatic effect on the light curve's shape; 
it simply provides an overall increase in its magnitude, and would be difficult to 
distinguish from amplification due to the galaxy hosting the microlens.

The interesting cases described above occurred because the supernova exploded
within the Einstein ring of a deflector, while the deflector proceeded to
move on a time scale comparable to the life of the supernova. \citet{Schneider87} 
found, in some instances, similar double peak modifications 
to the light curve caused by the supernova's photosphere
expanding into a deflector's critical point. 

The standard optical depth $\tau $ significantly
overestimates the probability that the above interesting cases will
occur. Using $\S$ 3 we can correct for the overestimate. For these
lensing configurations, the normalized impact parameter $u_{A}$ does not
exceed 0.1 ($A\approx 9$) which, using equation (\ref{p(A)}), 
gives a maximum probability
of $\sim 10^{-4}$ for $z_{s}\geqslant 1$ ($\tau =0.01$), and $\sim 10^{-3}$
for $z_{s}=2$ ($\tau =0.1$) if \citet{Zakharov04} are correct.
These numbers are somewhat higher than \citet{Linder88}
were predicting for Type I SN but not for Type II.
With a supernova rate of $R_{SNe Ia}= 2,000$ yr$^{-1}$   
the SN lensing rate at redshift $z=1$ is $\sim\, 10^{-4}\Omega$ yr$^{-1}$
\citep{Oguri03}.
Rates likes this, together with time scales of some cases studied here, imply that
such effects may not be observed unless a large number of cosmologically distant
(around $10^{4}$ for $z_{s}\geqslant 1$) type Ia supernovae are followed 
for a period of up to 2 years. For the high amplification cases with a second peak,
a seperate probability estimate can be made (see the Appendix).

\section{Conclusion}

We have shown that microlensing can significantly affect the light
curve of a cosmologically distant type Ia supernova. We restricted
our calculation to $z_{s}=1$ and $z_{d}=0.05$ in the currently accepted
$\Omega _{m}=0.3$, $\Omega _{\Lambda }=0.7$ flat cosmological model.
We found that microlensing not only increases the magnitude of the
light curve but also can cause a change in its shape. 
Relative transverse motion of the SN and 
lens, as well as the expanding photosphere \citep{Schneider87} 
can result 
in features such as a narrow but high peak, a plateau following the
peak, or even the presence of a second peak.

In $\S$ 4 we concluded that for microlensing
by compact masses distributed through the cosmos, the optical depth
is 0.01 $\left(z_{s}\sim 1\right)$ but might reach 0.1 $\left(z_{s}\sim 2\right)$,
meaning that the overall chance of a distant supernova type Ia being
microlensed is not negligible. Any multi-band supernova survey aimed
at finding supernovae at redshifts around $z=1$ (and above), could
discover and identify \emph{one} microlensed SN Ia event out of roughly
\emph{a hundred} events. However, the low impact parameters required to produce the 
special features depicted in
$\S$ 4 demand observation of $\sim 10^{4}$ supernovae at 
$z_{s}\geqslant 1$.
And, to see unusual features such as double peaks, the lensed supernova
must be followed for an extended period of $\sim$ 2 years. 
In the Appendix we have made an optical depth type estimate to include 
double peak events. For microlensing by stars in the bulge of a galaxy 
at $z_d=0.05$ we find a max probability of $\sim 1.7\times 10^{-3}$.
This is an ideal deflector distance for observing double peaks due to transverse motion. 
As expected this number is only slightly smaller than the 27\% optical
depth estimate made in $\S 3.3$ for bulge lensing when corrected for an impact 
$u=0.1$
These estimates
do not take into account the observational bias in favor of amplified
events ($\S$ 4) nor the possibly enhanced probability due to evolution 
of the rate at which Type-Ia supernovae have
occurred. It is interesting to note that according to
OGLE III \citep{Udalski03}
a histogram of $u$ values for Bulge lensing peaks at $u\sim 0.1$ (probably 
due to amplification biasing).  

\acknowledgements{}

The authors are pleased to thank an anonymous referee for helpful
criticisms of the first version of the manuscript and Zach Blankenship 
for correcting an oversight in this version. This work was in part 
supported by NSF grant AST0204771 and NASA grant NNG04GD36G.

\appendix
\section{Appendix}
In this appendix we compute the probability that a source,
followed for a period $T$, impacts a point deflector with a reduced
impact parameter less than $u\equiv b/R_E$ and simultaneously moves 
at least a distance $b$ during the period $T$. Such a time 
dependent impact will cause a change in the amplification of 
10\%-50\% depending on the actual impact. The idea here is to estimate
the chance of seeing a distortion in the light curve of a SN whose life 
time is $T_{SN}\sim 200$ days.

We start with a number density $N_d$ of  mass $m$ deflectors (located at a distance $D_d$ 
from the observer) moving with 
relative transverse velocities distributed according to:
\begin{equation}
\frac{dN_d}{dv}=N_d(D_d)\frac{v}{v_{rms}^2}e^{-v^2/2v_{rms}^2}.
\end{equation}
The probability of one of these moving deflectors impacting the line of 
sight to a source at $D_s$  with a reduced 
impact parameter $\leqslant u$ and moving a reduced distance $\geqslant u$ during 
a period $T$ is:
\begin{equation}
\Delta Prob(u,T) =\int^{D_s}_0dD_d\int^{\infty}_{u\,R_E/T}dvN_d(D_d)(\pi u^2R_E^2+2uR_Ev\,T)
\frac{v}{v_{rms}^2}e^{-v^2/2v_{rms}^2}.
\end{equation}
The integrand is the sum over areas represented in Figure 18.
The velocity integral can be done easily, 
and if the deflectors are effectively confined to a plane, 
the result can be written as

\begin{equation}
\Delta Prob(u,\xi) = \frac{4G}{c^2}\ \Sigma\,D\, u^2\left\{(\pi+2)
e^{-\xi^2/2}+\sqrt{2\pi}\frac{Erfc(\xi/\sqrt{2})}{\xi}\right\},
\end{equation}
where
\begin{equation}
\xi\equiv \frac{uR_E}{v_{rms}T}= u\,\frac{T_{rms}}{T},
\end{equation}
$\Sigma$ is the projected surface mass density
\begin{equation}
\Sigma \equiv \int^{D_s}_0m\,N_d(D_d)dD_d,
\end{equation}
and $Erfc$ is an error function.

The characteristic crossing time for microlensing is defined by 
$T_{rms}=R_E/v_{rms}=\sqrt{2r_S D}/v_{rms}$ (see $\S$ 3 for definitions)
which for Galaxy bulge-bulge lensing is about 10 days \citep{Udalski03}. 
For a similar 
galaxy at redshift $z=0.05$ 
lensing a distant SN through its bulge, the reduced distance $D$ 
is increased by a factor of $\sim 2.7\times 10^5$ [see $\S$ 3.3 and 
\citet{Han03}] and hence 
$T_{rms}$ increases to $\sim$ 5,200 days. If $u\sim 0.05$ and 
$T=T_{SN}\sim 200$ days, then $\xi \sim$ 1.3
and
$\Delta Prob(0.05,1.3)\sim 1.7\times 10^{-3}$. This particular probability falls 
off by at least an order of magnitude when $u<0.01$ or $u>0.15$.

\clearpage

\figcaption{} 
Model light curve of a supernova type Ia in its rest
frame. Observed data of SN 1990N are added to show the fit. 
\label{f1}

\figcaption{}
Microlensing by a point deflector. The deflector's
position is considered as the coordinate origin in the deflector plane.
\label{f2}

\figcaption{}
The position of the primary and secondary images
on the deflector plane (angular coordinates) with respect to source and
Einstein ring. Note that source makes an angle of 0.04 micro arcsecond
with respect to the deflector. 
\label{f3}

\figcaption{}
Amplification curves of images of an extended source
($z_{s}=1.0$) by a deflector at redshift $z_{d}=0.5$. The horizontal
axis is the angular separation of deflector and source, $\beta $,
(see Fig. 3) in micro arcsecond. 
\label{f4}

\figcaption{} 
Amplification curves of an extended source with a
radius of 178 AU, moving at 0.5 AU day$^{-1}$ in the deflector plane. The source
reaches its closest approach (impact parameter) at $t=0$. The figure
shows five trajectories with different impact parameters, b (see Fig. 8).
For this figure only, the surface brightness and radius are assumed constant. 
\label{f5}

\figcaption{} 
Normalized IDF profile calculated for SN\,Ia in U and
B bands using W7 model (courtesy of Eric Lentz, University
of Georgia), used to calculate the amplification curves of a supernova
as an extended source with radius-dependent surface brightness. In
this figure, a maximum expansion velocity of 30,000 km s$^{-1}$ has been
used to obtain the expansion velocity projected on the sky. The intensity 
profile used in the paper is also presented for photospheric speeds of 
13,000 km s$^{-1}$ and 15,000 km s$^{-1}$.
\label{f6}

\figcaption{} 
Expansion speed (\emph{upper panel}) and the radius (\emph{lower panel})
of the photosphere of a type Ia supernova as a function of time.
\label{f7}

\figcaption{}
Geometry of the microlensing model used in $\S$ 4. 
\label{f8}

\figcaption{}
Alification curve (\emph{upper panel}) and light
curve (\emph{lower panel}) of a supernova at $z_{s}=1.0$, microlensed
by a deflector at $z_{d}=0.05$. The input parameters are $m_{d}=1$ M$_{\odot },\,  
v=0.5$ AU day$^{-1}$, $b=0.0\, r_{E}$, and $l_{o}=0.0\, r_{E}$. 
\label{f9}

\figcaption{}
Same as Fig. 9, with input parameters $m_{d}=10^{-3}$ M$_{\odot },\,  
v=0.1$ AU day$^{-1}$, $b=0.1\, r_{E}$, and $l_{o}=0.25\, r_{E}$. 
\label{f10}

\figcaption{}
Same as Fig. 9, with input parameters $m_{d}=1$ M$_{\odot },\,  
v=0.5$ AU day$^{-1}$, $b=0.01\, r_{E}$, and $l_{o}=0.05\, r_{E}$. 
\label{f11}

\figcaption{}
Same as Fig. 9, with input parameters $m_{d}=10^{-3}$ M$_{\odot },\,  
v=0.1$ AU day$^{-1}$, $b=0.1\, r_{E}$, and $l_{o}=1.0\, r_{E}$. 
\label{f12}

\figcaption{}
Same as Fig. 9, with input parameters $m_{d}=10^{-3}$ M$_{\odot },\,  
v=0.5$ AU day$^{-1}$, $b=0.01\, r_{E}$, and $l_{o}=1.5\, r_{E}$. 
\label{f13}

\figcaption{}
Same as Fig. 9, with input parameters $m_{d}=10^{-3}$ M$_{\odot },\,  
v=0.5$ AU day$^{-1}$, $b=0.1\, r_{E}$, and $l_{o}=4.0\, r_{E}$. 
\label{f14}

\figcaption{}
Same as Fig. 9, with input parameters $m_{d}=1$ M$_{\odot },\,  
v=0.5$ AU day$^{-1}$, $b=0.0\, r_{E}$, and $l_{o}=0.1\, r_{E}$. 
\label{f15}

\figcaption{}
Same as Fig. 9, with input parameters $m_{d}=1$ M$_{\odot },\,  
v=0.5$ AU day$^{-1}$, $b=0.01\, r_{E}$, and $l_{o}=0.1\, r_{E}$. 
\label{f16}

\figcaption{}
Same as Fig. 9, with input parameters $m_{d}=10$ M$_{\odot },\,  
v=0.5$ AU day$^{-1}$, $b=0.0\, r_{E}$, and $l_{o}=0.05\, r_{E}$. 
\label{f17}

\figcaption{}
Area on the deflector plane within which a microlens 
would be close enough to a luminous SNe Ia to  cause significant changes in 
its lightcurve.

\label{f18}

\clearpage

\epsscale{.8}

\plotone{f1.eps}
\begin{center} Figure 1 \end{center}
\eject

\plotone{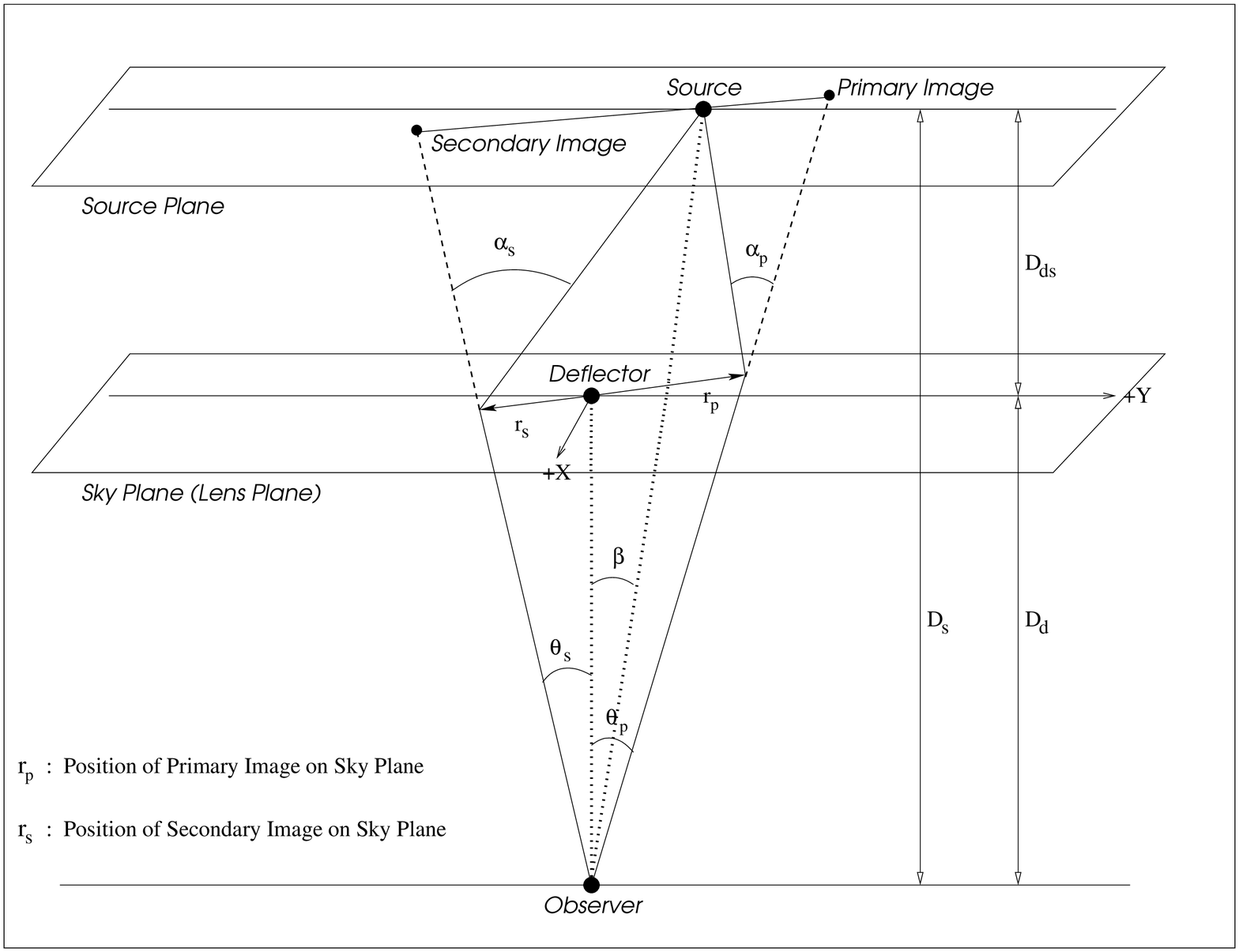}
\begin{center} Figure 2 \end{center}
\eject

\plotone{f3.eps}
\begin{center} Figure 3 \end{center}
\eject

\plotone{f4.eps}
\begin{center} Figure 4 \end{center}
\eject

\plotone{f5.eps}
\begin{center} Figure 5 \end{center}
\eject

\plotone{f6.eps}
\begin{center} Figure 6 \end{center}
\eject

\plotone{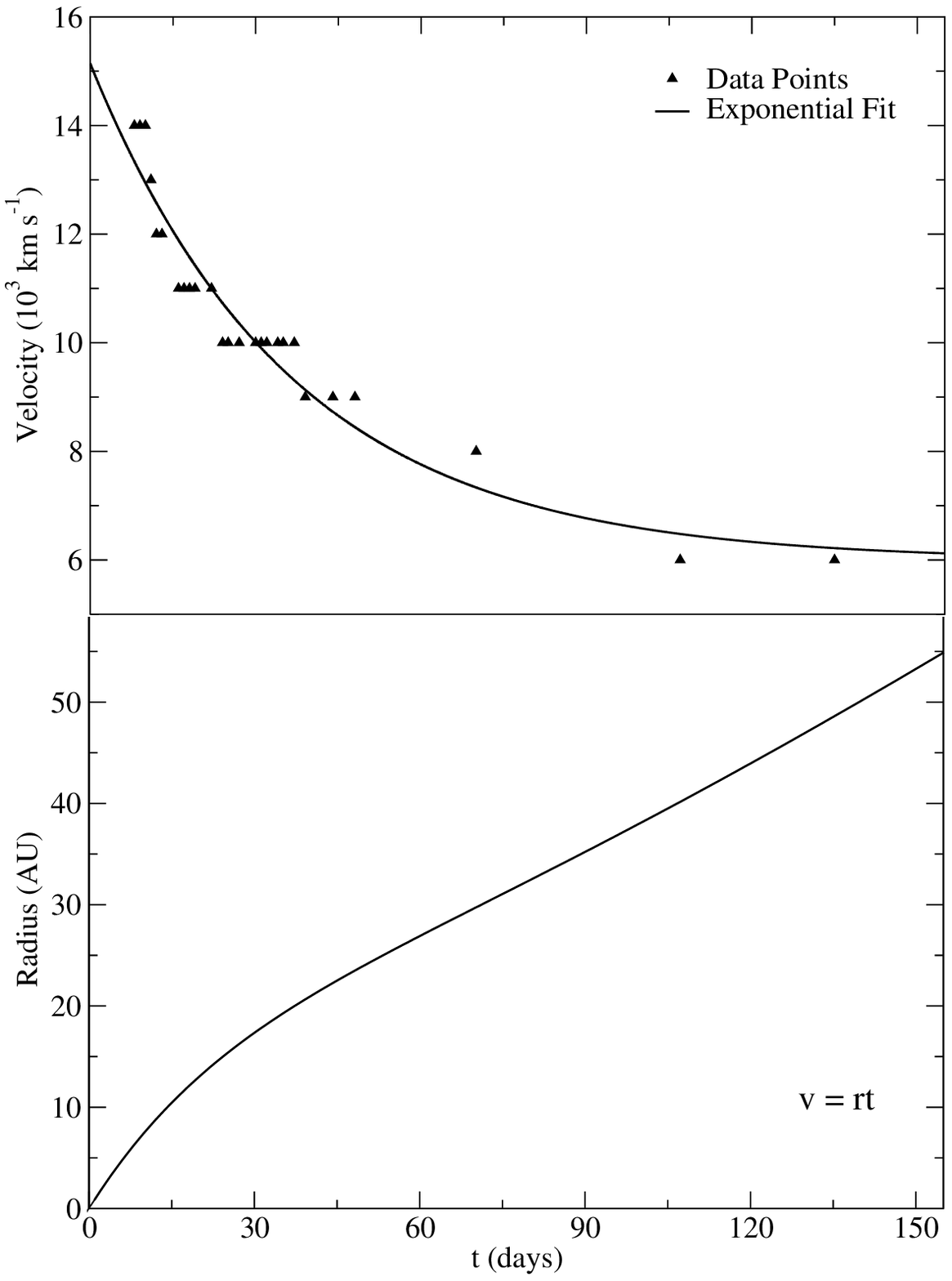}
\begin{center} Figure 7 \end{center}
\eject

\plotone{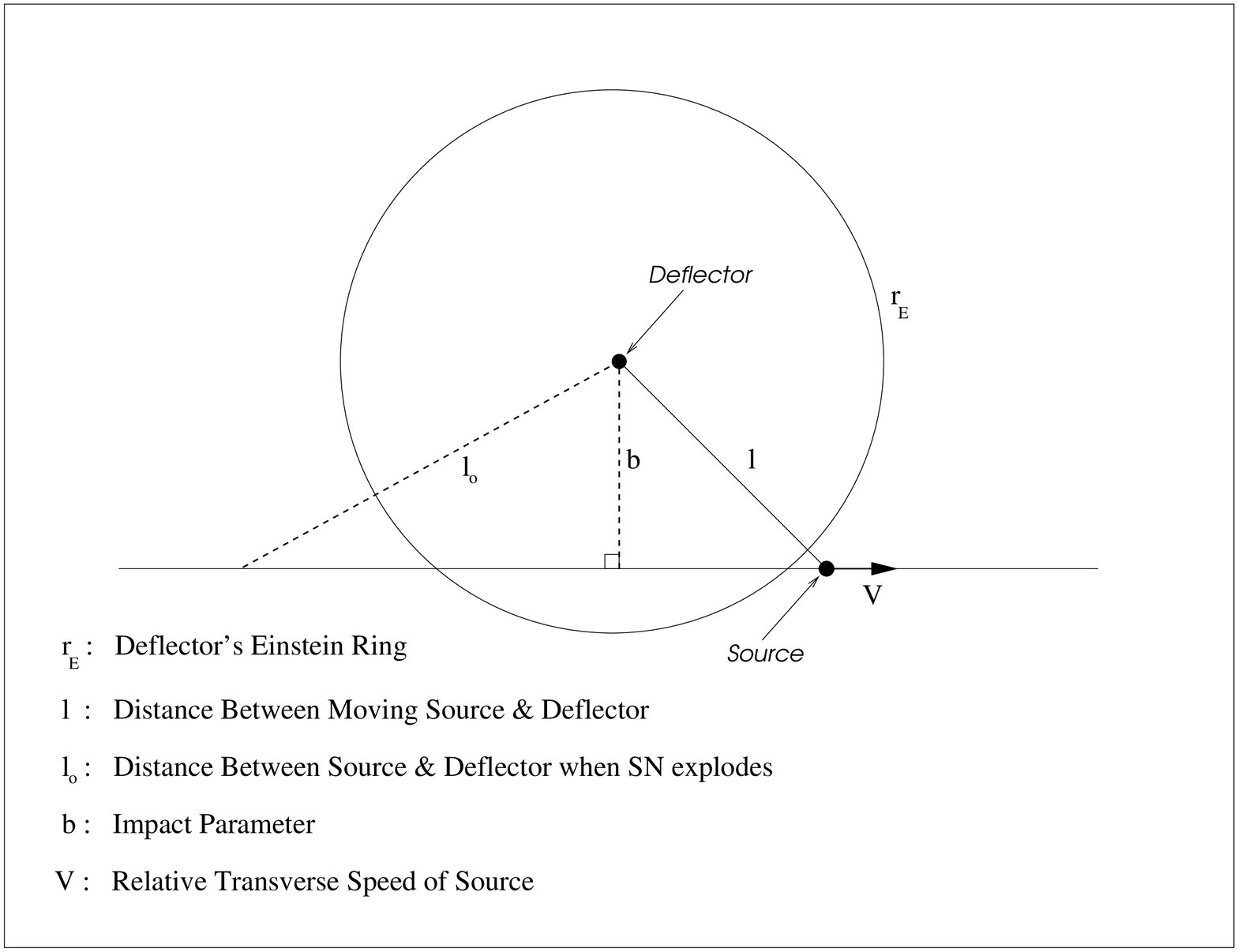}
\begin{center} Figure 8 \end{center}
\eject

\plotone{f9.eps}
\begin{center} Figure 9 \end{center}
\eject

\plotone{f10.eps}
\begin{center} Figure 10 \end{center}
\eject

\plotone{f11.eps}
\begin{center} Figure 11 \end{center}
\eject

\plotone{f12.eps}
\begin{center} Figure 12 \end{center}
\eject

\plotone{f13.eps}
\begin{center} Figure 13 \end{center}
\eject

\plotone{f14.eps}
\begin{center} Figure 14 \end{center}
\eject

\plotone{f15.eps}
\begin{center} Figure 15 \end{center}
\eject

\plotone{f16.eps}
\begin{center} Figure 16 \end{center}
\eject

\plotone{f17.eps}
\begin{center} Figure 17 \end{center}
\eject

\plotone{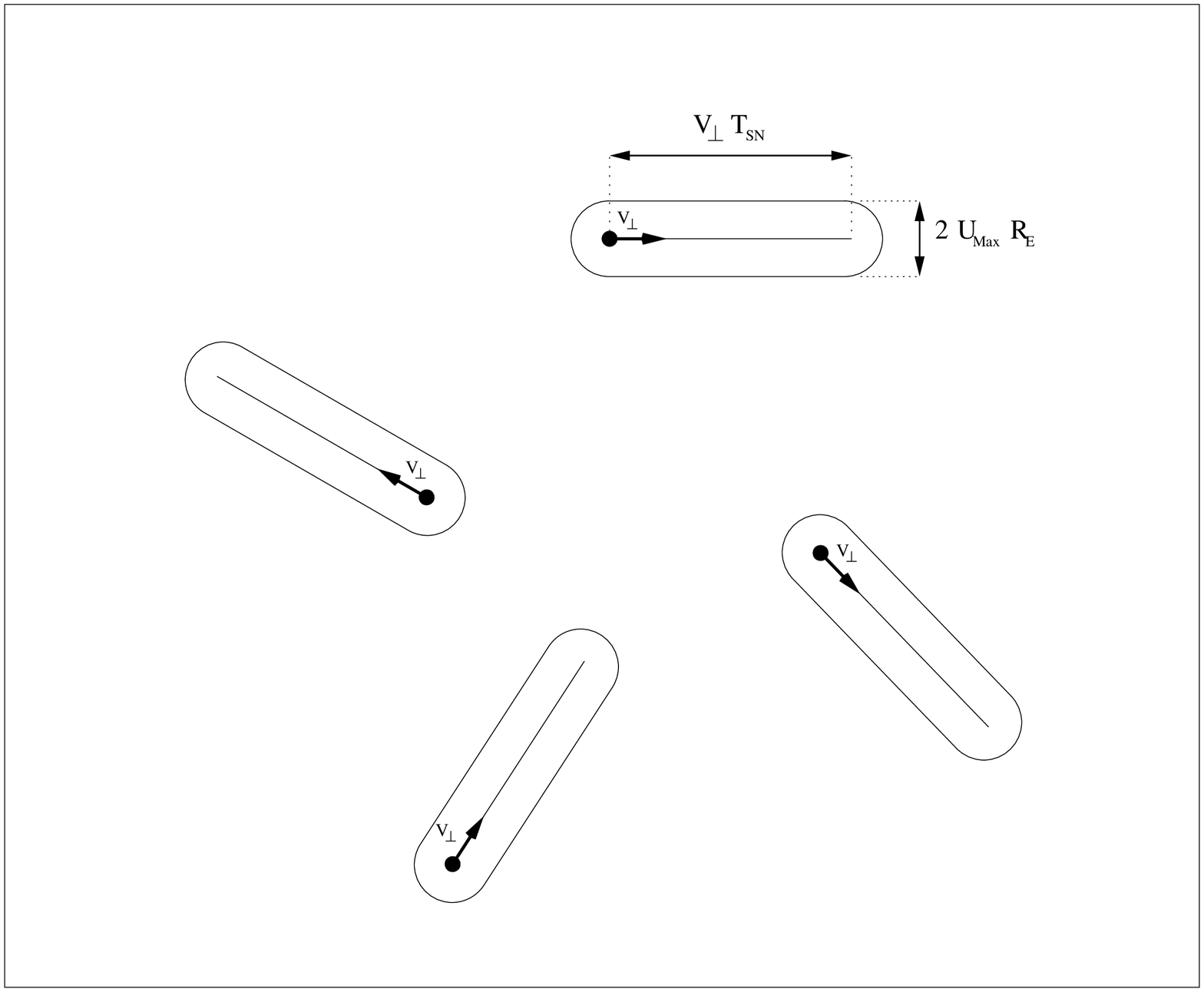}
\begin{center} Figure 18 \end{center}
\eject

\end{document}